\documentclass[aps,12pt,nofootinbib]{revtex4}

\usepackage{graphicx,graphics}
\usepackage{enumerate}
\usepackage{subfigure}
\usepackage{amsmath}
\usepackage{amsfonts}
\usepackage{amssymb}
\usepackage{amsthm}
\usepackage{psfrag}

\catcode`\@=11
\def\numberbysection{\@addtoreset{equation}{section}
\def\theequation{\arabic{section}.\arabic{equation}}}
\numberbysection

\begin{document}

\title{Rotor-Router Walk on a Semi-infinite Cylinder}

\author{Vl.V. Papoyan$^{1, 2}$, V.S. Poghosyan$^3$ and V.B. Priezzhev$^1$}
\affiliation{
$^1$Bogoliubov Laboratory of Theoretical Physics,\\ Joint Institute for Nuclear Research, 141980 Dubna, Russian Federation\\
$^2$Dubna State University, Dubna, Russian Federation\\
$^3$Institute for Informatics and Automation Problems\\ NAS RA, 0014 Yerevan, Republic of Armenia
}

\begin{abstract}
We study the rotor-router walk with the clockwise ordering of outgoing edges on the semi-infinite cylinder. Imposing uniform conditions on the boundary of the cylinder, we consider growth of the cluster of visited sites and its internal structure. The average width of the surface region of the cluster evolves with time to the stationary value by a scaling law whose parameters are close to the standard KPZ exponents. We introduce characteristic labels corresponding to closed clockwise contours formed by rotors and show that the sequence of labels has in average an ordered helix structure.
\end{abstract}

\maketitle

\noindent \emph{Keywords}: rotor-router walk, roughening phenomena, helix structure.

\section{Introduction}
A question raised in one of the first papers on the rotor-router walk (introduced originally under the name ``Eulerian walkers model'' \cite{PDDK})    concerned  the shape and size of the cluster of sites visited by the walker till time $t$. The finite-time behavior of the rotor walk on an infinite lattice is a much more difficult problem than the recurrent rotor walk on a finite graph which admits a theoretical treatment due to the Abelian structure similar to that in the sandpile model \cite{Dhar} (see \cite{HLMPPW} for more references).
Heuristic arguments used in \cite{PDDK} showed that, given a uniform random initial configuration of rotors on the infinite square lattice,
the average cluster of visited sites is roughly a disc of radius $R$ which grows as $t^{1/3}$.

The problem of circular shape appears also in the model of rotor aggregation introduced by Propp \cite{LP05,LP07}.
In this model, $n$ particles starting at the origin perform rotor-router walk until reaching an unoccupied lattice site which then becomes occupied.
Levine and Peres \cite{LP07} proved that for $d$-dimensional lattice the asymptotic shape of the set of occupied sites is an Euclidean ball.

The shape of the cluster of sites visited by a single rotor walker is another problem. In contrast to the rotor aggregation where each particle finishes its trajectory at the first unoccupied site, the single rotor walk continues its motion permanently modifying the rotor configuration. Then, an additional
essential problem appears on the number of returns of the rotor walk to the origin.

Recently, Florescu, Levine and Peres \cite{FLP} have proved that the number of distinct sites visited by the rotor-router walk on $d$-dimensional lattice in $t$ steps is greater or equal $c t^{d/(d+1)}$, where $c$ is a positive constant.
Particularly for $d=2$, this statement gives a tight lower-bound of the law $t^{1/3}$ for the growth of the average linear size of a two-dimensional cluster, but leaves aside the question of its asymptotic form.

Kapri and Dhar \cite{Kapri} used extensive numerical simulations to study the shape of the rotor-router cluster  formed by an Eulerian walk on a random background in two dimensions and presented evidence that it converges to a perfect circle for large number of steps.
Further attempts to describe the structure of the cluster of visited sites were made in \cite{PPP1,PPP2}.
The motion of the rotor-router walker creates a sequence of loops of rotors and a set of labeled sites where the loops become closed.
In the case when the rule of clockwise rotations is applied to all rotors, the sequence of labeled sites forms a spiral-like structure which was conjectured to tend asymptotically to the {\it Archimedean spiral} in average \cite{PPP1,PPP2}.
The Archimedean property is consistent with the perfect circular asymptotic shape of the cluster of visited sites.
Also, it agrees with a scaling law for the number of visits to a site separated from the origin by distance $x$ for $N$ steps of the walker \cite{Kapri}.

While the circular shape of the rotor-router cluster is approved at least numerically, the fluctuations of this shape are more resistant to a clear description.
A standard approach to investigation of fluctuations of a growing surface implies a formulation in terms of the KPZ theory \cite{KPZ}.
Kapri and Dhar \cite{Kapri} calculated the width $W$ of the surface region of the rotor-router cluster on the square lattice to determine the saturation-width exponent $\alpha$ and the dynamical exponent $z$ in the KPZ scaling law
\begin{equation}
W\sim L^{\alpha}f_{KPZ}(t/L^z),
\label{KPZ}
\end{equation}
where $L$ is a characteristic length of the surface, and $f_{KPZ}(x)$ is the scaling function which behaves as $x^{\alpha
/z}$ for $0 < x \ll 1$ and tends to 1 for $x \gg 1$ \cite{KPZ}. Since for the growing two-dimensional cluster of radius $R$, both time $t$ and length $L$ are proportional to $R$, the expected KPZ values $\alpha=1/2$ and $z=3/2$ lead to asymptotic law
$W\sim R^{\gamma}=R^{1/3}$.

The exponent $\gamma$ obtained in \cite{Kapri} was estimated as $\gamma=0.40\pm0.06$. This value of $\gamma$ is consistent with the KPZ exponents although it was difficult to control its limiting behavior. A non-controlling influence of shape fluctuations can be seen also in a slow convergence of the spiral structure to the proposed Archimedean law \cite{PPP2}. A possible reason for these difficulties is a lack of the stationary state for fluctuations of the boundary of the radially growing cluster and an unrestricted growth of the boundary region $W$ which stands in contrast to an expected stationary regime for the infinitely long cylinder. Singha \cite{singha} compared surface fluctuations in growing clusters  on the plane and those in the cylindrical geometry where width $W$ tends to a constant. The comparison showed quite different behavior of fluctuations in two geometries: the auto-correlation function tends with time to a non-zero value for the plane and decays to zero for the long cylinder.

An obvious advantage of the cylindrical geometry is the explicit separation of variables $L$ and $t$ in the scaling law (\ref{KPZ}) instead of their merging to single variable $R$ in the plane geometry. This allows to investigate the asymptotic behavior for large $L$ and large $t$ separately. As to detailed properties of the growing clusters, we may expect more regular behavior of the spiral structure which should be converted into a regular helix in the cylindrical case.

In this paper, we consider the rotor-router walk on a semi-infinite cylinder of square lattice with a boundary of perimeter $L$.
Rotors are considered as arrows attached to each lattice site and directed toward one of four neighbors on the lattice.
Arrows at the boundary sites have three possible directions.
A particle called usually {\it chip}, performs a walk jumping from a site to its neighbor along the current direction of rotor in the site.
Arriving to a given site, the chip rotates the arrow 90 degrees clockwise and moves toward the site pointed by new direction of the arrow.
On the boundary, the arrow pointed clockwise with respect to the top of the cylinder turns 180 degrees and becomes pointed anticlockwise.

Imposing special initial directions of rotors on $L$ boundary sites and random initial directions in the remaining lattice we provide a uniform propagation of the cluster of visited sites downward from the top of the vertically oriented semi-infinite cylinder.
Then the asymptotic average  boundary of the growing cluster is flat and the non-trivial shape problem vanishes.
The average width $W$ of the surface region evolves with time to its stationary value presumably by the KPZ law (\ref{KPZ}).
A check of the scaling form of $W(L,t)$ and determination of the related exponents is the first goal of the present paper.

Another goal is an analysis of the internal structure of the growing clusters. There are plenty of models demonstrating the local growth mechanism and kinetic roughening phenomena, for instance, random and ballistic deposition, the Eden model, RSOS models and polymer growth (see \cite{KPZ} for references).
The growth of the cluster of visited sites generated by the rotor walk has some peculiarities.
Florescu, Levine and Peres \cite{FLP} defined an {\it excursion} from the origin $o$ as the rotor walk started at $o$ and runs until it returns to $o$ exactly once from each of the neighboring sites.
The cluster appearing at the end of each excursion consists of rotors oriented towards the boundary of cylinder.
These rotors are directed edges of a graph which is a forest of trees attached to the boundary sites.
Each excursion adds to the forest additional branches.
The accumulation of edges is accompanied by a coarse-grained accumulation of domains formed by loops of rotors, in particular the clockwise closed contours.
The latter process imposes a larger scale of the fluctuations of growing surface and is responsible for the asymptotic values of the sought exponents.

Beside the clusters of visited sites, we consider spiral structures introduced in previous papers \cite{PPP1,PPP2}.
The site, where the chip completes a next clockwise contour, was called {\it label}.
We have noticed in \cite{PPP1,PPP2} that the sequence of labels forms a spiral structure in the case of plane geometry. After averaging over initial random configurations of rotors, the sequence approaches asymptotically the  Archimedean spiral. In the cylinder geometry, the spiral should be converted into a cylindrical helix. The determination of its parameters and examination of
the convergence to the limiting form is the concluding part of the work.

\section{Definitions and basic theorems}

Let $G=(V,E)$ be an infinite directed graph whose vertices $v \in V$ are sites of a semi-infinite cylinder of square lattice with a boundary $B \subset V$ of perimeter $L$. Let $E_v$ be the set of outgoing edges of site $v$. It consists of four outgoing edges $E_v^0,E_v^1,E_v^2,E_v^3$ for bulk sites $v \in V\setminus B$ and of three outgoing edges $E_v^0,E_v^1,E_v^2$ for boundary sites $v \in B$. In other words, $\deg(v)=4$ for the bulk sites and $\deg(v)=3$ for the boundary sites.
A rotor configuration is defined as a collection of outgoing edges $\rho(v)\in E_v$ from every site of the lattice. We assume that elements of both the bulk and the boundary sets $E_v$ are ordered locally clockwise.

A chip-and-rotor state $(w,\rho)$ is a pair consisting of a vertex $w \in V$ which represents the location of the chip and a rotor configuration $\rho(v)$.
The rotor-router walk is a sequence of chip-and-rotor states $(w_0,\rho_0)$,$(w_1,\rho_1)$,$(w_2,\rho_2),\dots $.
At each step $(w_t,\rho_t) \rightarrow (w_{t+1}, \rho_{t+1})$ of discrete time $t$, the chip arriving to site $w=w_t$, changes the position of rotor $\rho(w)=E^i_w$ to $E^j_w $ where index $j$ is changed as $j=i+1(\mod 4)$ if the vertex $w$ is in bulk and $j=i+1(\mod 3)$ if the vertex $w$ is on the boundary. Then, the chip moves to the neighboring site of $w$ pointed by the new position of the rotor.

The motion of the chip is determined by the initial chip-and-rotor state $(w_0,\rho_0)$.
We denote boundary sites $v\in B$ by numbers $1,2,\dots,L$.
The initial rotor configuration $\rho_0(v)$ consists of specifically chosen boundary rotors $\rho_0(i)$, $i=1,2,\dots, L$ and the randomly chosen rotor configuration $\rho_0(v)$ for $v \in V\setminus B$.
A general aim is a characterization of rotor states $\rho_t$ for $t>0$ and description of clusters of sites visited by the rotor walk.

Given a rotor state, we say that a group of rotors outgoing from sites  $v_1,v_2, \dots, v_{n+1}$ forms a directed path if  $v_i$ and $v_{i+1}$ are neighbors for
all $i=1,\dots,n$ and the rotor at $v_i$ is directed toward $v_{i+1}$. The directed path of rotors becomes a cycle if  $v_1 = v_{n+1}$.
A shortest possible cycle consists of two adjacent sites $v_1$, $v_2$, which are connected by a pair of edges from $v_1$ to $v_2$ and back. We call such cycles {\it dimers} by analogy with lattice dimers covering two neighboring sites. A cycle formed by more than two edges is called {\it contour}. If rotors belonging the contour are oriented clockwise, we call it the clockwise contour. The clockwise or anticlockwise directions depend on an orientation of the surface. Below, we will fix the orientation of the surface by the mapping of the cylinder onto the two-dimensional annulus.

We use two technical tools for analysis of the structure of growing clusters of visited sites. They are: a selection of clockwise contours among the rotor configurations \cite{PPP1}, and a representation of the recurrent rotor-router walk as a sequence of excursions \cite{FLP}. A special role of clockwise contours follows from a property of the chip-and-rotor states proved in \cite{PPP1}  and called {\it week reversibility}. A precise formulation of this property is given as Theorem 2 in \cite{PPP1}. Here we need a reduced form of Theorem 2 and its corollary, which can be formulated as follows.

{\it Proposition 1}. Let $C$ be the clockwise contour on a planar graph consisting of vertices $w_1,w_2,\dots,w_n$ ordered anticlockwise and containing an arbitrary rotor configuration inside allowing a recurrent rotor walk. The rotor-router operation is applied to the chip at $w_1\in C$ and continues until the moment when the chip returns to $w_1$ and the rotor $\rho(w_1)$ on the contour is made oriented anticlockwise. Then all rotors on $C$ are becoming oriented anticlockwise and the moments $t_1,t_2,\dots,t_n$ when the rotors at $w_1,w_2,\dots,w_n$ become anticlockwise for the first time are ordered as $t_1 < t_{2} < \dots  < t_n$.

{\it Remark}. The semi-infinite cylinder can be considered topologically as a planar annulus. Then, the recurrence of the rotor configurations in the Proposition 1 guarantees that any rotor walk started at a contour enveloping the cylinder returns to the contour.

The proof of the first part of the Proposition 1 is shared with that of Theorem 2 in \cite{PPP1} while the second part is proved as Corollary  \cite{PPP1}.
\begin{figure}[!ht]
\includegraphics[width=130mm]{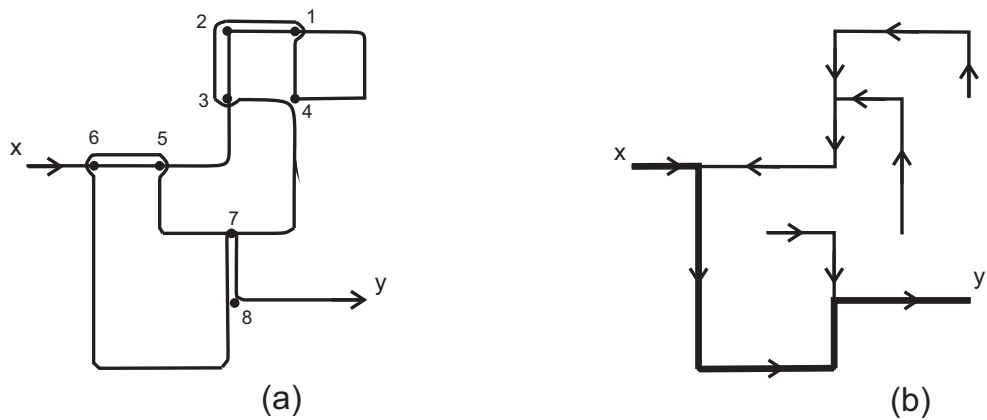}
\vspace{-1mm}
\caption{\label{loops} (a) The chip trajectory between vertices $x$ and $y$. The closed circles denote vertices where cycles become closed.
(b) The resulting configuration of rotors at the moment when the chip reaches vertex $y$. The bold line denotes a backbone of the resulting tree.
Arrows show orientations of branches formed by rotors.}
\end{figure}

Figure \ref{loops} shows how Proposition 1 works. Consider a chip trajectory between vertices $x$ and $y$ in Fig.\ref{loops}(a). At moment $t_1$, the chip visits the vertex denoted by 1 where the first clockwise contour becomes closed. The chip enters the contour, reverses its orientation into anticlockwise by Theorem 1 and leaves vertex 1 in left direction at moment $\bar{t}_1$. Due to Proposition 1, we can replace the steps between $t_1$ and  $\bar{t}_1$ by the single operation of reversion of orientation of the first contour. The next two vertices where closed cycles appear are 2 and 3. The both corresponding cycles are dimers. Leaving dimers, the chip forms the anticlockwise contour of vertices 1,2,3,4 and
leaves it in vertex 4 at the next step. Then, the trajectory continues up to vertex 5 where the second clockwise contour becomes closed at moment $t_2$. The chip leaves this contour at $\bar{t}_2$ and we again skip the steps between $t_2$ and $\bar{t}_2$ reversing the orientation of the contour. Continuing, the chip closes the dimer cycle in vertex 6, the anticlockwise contour in 7, the dimer cycle in 8 and finally reaches vertex $y$.
The configuration of rotors obtained as a result of the chip motion from $x$ to $y$ is shown in Fig.\ref{loops}(b). Since all cycles were opened
during the rotor walk, the resulting configuration is a tree. Vertices $x$ and $y$  appear to be connected by a line which can be considered as a backbone of the tree. Thus, a complicated trajectory of the rotor-router walk can be represented as a directed path and a collection of tree branches attached to the path.

The backbone of the graph generated by the rotor-router walk can be compared with the Loop Erased Random Walk (LERW) introduced by Lawler \cite{Lawler}.
Like the LERW, the rotor backbone avoids closed loop and, after sufficiently large number of steps, can be considered as a "chemical path" of the resulting spanning tree. However, the spanning tree whose chemical path corresponds to the LERW is so called {\it uniform} spanning tree \cite{Wilson} and this chemical path has the fractal dimension $\nu_{LERW}=5/4$ \cite{Con,Satya}. We will see below that the spanning tree generated by the single rotor walk in a finite volume is not uniform. For a large volume, the fractal dimension of its chemical path $\nu_{rotor}=1$ as it follows from numerical simulations. A rigorous proof of  $\nu_{rotor}=1$ is still absent.

The second tool we use below is the decomposition of the rotor walk into excursions.
The definition of excursions given in \cite{FLP} reads: Fix a vertex $o\in V$. An excursion from $o$ is a rotor walk started at $o$ and run until it returns to $o$ exactly $\deg (o)$ times. Lemma 2.4 in \cite{FLP} states the following properties of excursions.

(i)\ If time of $(n+1)$-th excursion is finite, the number of visits of site $x$ during this excursion does not exceed $\deg(x)$ for all $x\in V$.

(ii)\ Let $A_n$ be the set of sites visited during $n$-th excursion. Then  the number of visits of site $x$ during $(n+1)$-th excursion is $\deg(x)$ for all $x\in A_n$.

(iii)\  $A_{n+1}\supseteq A_n \cup \partial A_n$ where $\partial A_n$ is the set of all vertices neighboring to $A_n$ and not belonging to $A_n$ .

Since each excursion is a closed path, the backbone either collapses into a single point, if the backbone path on the lattice is contractible,
or becomes a non-contractible loop enveloping the cylinder in the case of cylindrical geometry. In what follows, we consider the latter case imposing a specific initial conditions on the top boundary of the cylinder.

\section{Rotor configurations and the cluster growth }

\begin{figure}[!ht]
\includegraphics[width=65mm]{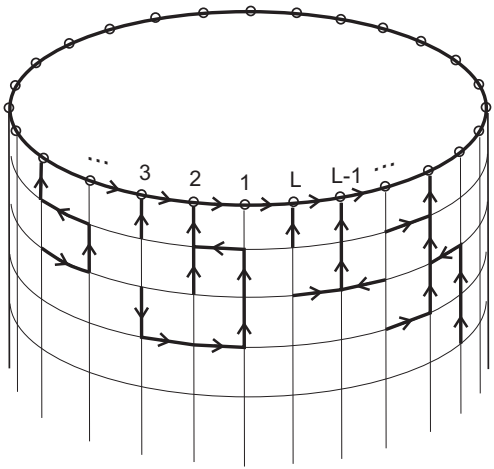}
\vspace{1mm}
\caption{\label{Cyl} Semi-infinite cylinder. Arrows at boundary sites denote rotors directed anticlockwise in the initial rotor configuration.
Arrows at bulk sites denote rotors constituting trees attached to the boundary. }
\end{figure}

Consider the initial rotor configuration $\rho_0(v)$ shown in Fig.\ref{Cyl}. All boundary rotors  $\rho_0(i)$, $i=1,2,\dots, L$ are oriented
anticlockwise. The bulk rotors are chosen randomly with equal probabilities for four possible directions $E_v^0,E_v^1,E_v^2,E_v^3$. The bulk rotors near boundary can be combined to form a forest of oriented trees attached by their roots to the boundary. If there are no rotors directed to a given boundary vertex, the tree consists of the root alone.
We put the initial position of the chip $w_0$ to site $1$ and consider evolution of the chip-and-rotor state $(w_0,\rho_0)$,$(w_1,\rho_1)$,$(w_2,\rho_2),\dots $ in discrete time.
The rotor walk is called {\it recurrent} if it visits the initial vertex infinitely many times, otherwise it is {\it transient} \cite{AngelHol,HussSava}. If the initial configuration $\rho_0(v)$ has an infinite path to initial vertex 1, then the walk is transient \cite{FLP}. Below, we need the recurrence property, then we assume that all random trees oriented to the boundary are finite, because otherwise they contain an infinite path.

The surface of the semi-infinite cylinder is equivalent topologically to a two-dimensional annulus whose internal ring coincides with the top boundary of the cylinder and the external ring is situated at infinite distance from the internal one. The anticlockwise direction of rotors in
Fig.\ref{Cyl} corresponds to the clockwise orientation of the internal ring against the surface of the annulus. The external ring is not reachable for the recurrent rotor walk, so its orientation does not matter. Then the initial rotor configuration $\rho_0(v)$ meets the conditions of Proposition 1 and we can assert after $n$ steps the existence of a chip-and-rotor state $(w_0,\rho_n)$ where all boundary rotors changed their orientation to opposite for the first time.

During the next $L$ steps $n+1,n+2,\dots,n+L$, the chip moving along the boundary rotates the boundary rotors at $1,L,L-1,\dots,2$ one by one so that at $(n+L)$-th step
the boundary rotors return to the initial positions and the chip is in the vertex 1. The sequence of steps from the first to $(n+L)$-th fits the definition of excursion given in \cite{FLP}. Denote the time when the first excursion is completed by $\tau_1$. Lemma 2.2 in \cite{FLP} claims that if $\tau_1$ is finite and there is
a directed path of initial rotors from vertex $x$ to the origin 1, then the rotor at $x$ performs the full rotation. Therefore, we come to an important conclusion: given a forest of trees attached to the boundary in the initial rotor configuration $\rho_0(v)$, the first excursion produces a new configuration $\rho_{\tau_1}(v)$ which contains the rotors of the initial forest at the same positions as in $\rho_0(v)$.

Proposition 1 allows us to say more about results of the first excursion. Consider a vertex $v\in V$ which is visited by the chip during the first excursion but does not belong to the initial boundary forest. Since the chip returns to the origin, there is a rotor configuration $\rho_t$ at step
$t$, $1<t<\tau_1$ where $v$ belongs to a contour $C_v$. If $C_v$ is clockwise, it will be open after reversing by Proposition 1; if $C_v$ is anticlockwise or it is a dimer, it will be open at next step $t+1$. In all cases $v$ will belong to a branch attached to the initial boundary forest. Therefore, during the first excursion the initial forest will be augmented by new branches containing all visited vertices not belonging to the initial forest. In particular, all vertices separated from the initial boundary trees by distance 1 are connected to them after the first excursion due to the rotor rotation rule.

The moments of time when new branches are attached to the trees rooted at $1,2,\dots,L$ are strictly ordered: no new branches are attached to $i$-th tree,
$1 <i\leq L$, until the replenishing of the $(i-1)$-th tree is completed. Indeed, the second part of Proposition 1 claims that the moments $t_1,t_2,\dots,t_L$ when the rotors at $1,2,\dots,L$ reverse their direction are ordered as $t_1<t_2<\dots,<t_L$. Any contour producing a branch of $(i-1)$-th tree should be closed before $t_{i-1}$ and a new branch can be added to $i$-th tree only after the moment when the chip enters the boundary site $i$ i.e. later than $t_{i-1}$.

The next excursions act similarly. Properties (ii) and (iii) of the excursions allow to consider each new excursion in the same way as the first one with a given initial rotor configuration. The rotor configuration $\rho_{\tau_n}$ obtained after $n$-th excursion contains the same (anticlockwise) rotors at boundary vertices $1,2,\dots,L$ and differs from $\rho_0(v)$ by the size of forest only. Each excursion adds new branches to the existing trees, so that the forest of trees attached to the boundary grows monotonically. If after some excursion a tree appears which has no free neighboring vertices not visited by the rotor walk during previous excursions, then the growth of the tree
stops. Actually, for a sufficiently large number of excursions and finite $L$, a single tree remains which continues its growth, whereas remaining $L-1$ trees are surrounded by branches of neighboring trees.
We can conclude that, given an arbitrary height $H$ of the part of cylinder, there is a number of excursions $n_H$ after which there are no free vertices with the vertical coordinate $h<H$, so that the obtained forest spans the lattice restricted by height $H$.

In two forthcoming sections, we consider statistical properties of a boundary of the forest and an internal structure of the cluster of sites visited by the rotor walk.

\section{ Statistical properties of the cluster boundary}

Starting with a recurrent initial configuration $\rho_0(v)$, the rotor walk generates a forest which covers all sites visited for time $t$. The surface of the forest is the set of visited sites which have at least one neighboring site not belonging to the forest.
For each vertex $v\in V$, we define the height $h(v)$ as a vertical distance of $v$ from the boundary of the cylinder.
The average height of the surface $H(t)$  at moment of time $t$ is the average of  $h(v)$ over all surface sites. The average number of visited sites is
proportional to $H(t)L$, so we have for the velocity of growth
\begin{equation}
\frac{dH(t)}{dt}\sim \frac{1}{H(t)L}
\label{growth}
\end{equation}
This implies that for large time $t$
\begin{equation}
H(t) \sim \left(\frac{t}{L}\right)^{1/2}
\label{Growth}
\end{equation}

The time variable $t$ needs some elaboration. The most natural representation of the discrete time is the number of steps $t\in {0,1,2,\dots}$ of the rotor-router walk. However, the evolution of surface $H(t)$ proceeds in a coarsened scale determined by the sequence of excursions. Then, a convenient variable is $t_n, n=1,2,\dots$, where $t_n$ is the moment when the $n$-th excursion is completed.

In Section II, we defined $A_n$ as the set of sites visited during the $n$-th excursion. The surface of the set of visited sites after $n$-th excursion is subset  $\Gamma(A_n) \subseteq A_n$  of visited sites that have at least one neighbor unvisited during first $n$ excursions  or, by property (iii), during
the last $n$-th excursion.

To check if the growth of the cluster of visited sites is within the KPZ class of universality, consider the average width of the surface $\Gamma(A_n)$, defined as
\begin{equation}
W(L,n)= \left\langle \frac{1}{\#\Gamma(A_n)} \sum_{v \in \Gamma(A_n)} \left( h(v) - \overline{h}\, \right)^2 \right\rangle^{1/2},
\label{W-KPZ}
\end{equation}
where
\begin{equation}
\overline{h} = \frac{1}{\#\Gamma(A_n)} \sum_{v \in \Gamma(A_n)} h(v).
\end{equation}
Numerical simulations for the cylinders with circumference $L=100,200,\ldots,1000$, number of excursions $ N_{ex} = n \leq 200$ and number of samples $= 10^4$ showed that

\begin{equation}
W(L,n) \sim L^\alpha f(t_n/L^z),
\end{equation}
with scaling function $f(u)$ satisfying

\begin{equation}
f(u) \sim \left\{
          \begin{array}{ll}
              u^\beta, & u \ll 1 \\
              1, & u \gg 1
          \end{array}
          \right. \quad .
\end{equation}
The estimated values of exponents are $\alpha = 0.51\pm 0.03$, $\beta =0.35 \pm 0.05$ in a good agreement with predictions of the KPZ theory $\alpha = 1/2$, $\beta =1/3$.
Average time $<t_n>$ when $n$-th excursion is completed is proportional to $n^2$ for sufficiently large $n$.

\section{The helix structure}
In section III, we have considered the evolution of the chip-and-rotor state  $(w_t,\rho_t)$ in terms of excursions. Now, we consider the evolution as a sequence of clockwise loops. The graph representation of state  $(w_t,\rho_t)$ is a spanning subgraph $G_s \subset G$ whose edges coincide with the current positions of rotors in $\rho_t$ and a selected vertex $w_t\in V$ shows the chip location. Each step of evolution $(w_t,\rho_t) \rightarrow (w_{t+1}, \rho_{t+1})$ moves the chip in one of neighboring lattice sites and modifies $G_s$ in vicinity of $w_t$. If at some moment of time $t=t_1>0$, the chip occurs on a clockwise loop for the first time, we say that the rotor walk creates a clockwise contour $C_1\subset G_s$ and label the vertex $w_{t_1}$ by $\alpha_1$. By Proposition 1, the subsequent evolution reverses clockwise contour $C_1$ into anticlockwise $\bar{C_1}$ and the chip leaves $\bar{C_1}$ at moment $\bar{t_1}$. We skip the part of evolution between $t_1$ and $\bar{t_1}$ and continue the rotor walk since $\bar{t_1}$. A next clockwise contour $C_2$ can be created at moment $t_2>\bar{t_1}$ outside $C_1$ (but can be adjacent with $C_1$ or may contain $C_1$ inside). Again,
we label the vertex $w_{t_2}$ by $\alpha_2$, skip the evolution between $t_2$ and $\bar{t_2}$ and continue till moment $t_3$ when $C_3$ appears.
As before, $C_3$ is outside $C_1$ and $C_2$ but can be adjacent with them or contain one of them or both $C_1$ and $C_2$ inside.
In this way, we obtain a sequence of distinct labels $\alpha_1,\alpha_2,\dots$ which mark the sequence of appeared clockwise contours.

\begin{figure}[!ht]
\includegraphics[width=120mm]{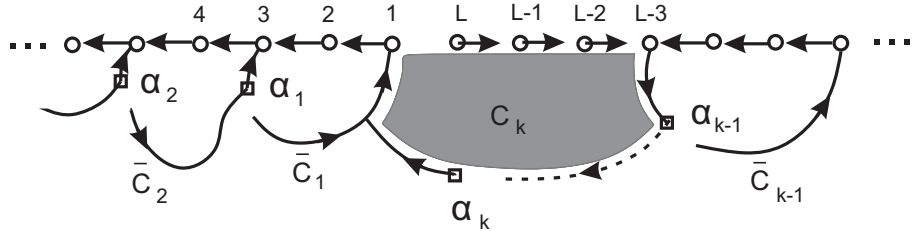}
\vspace{1mm}
\caption{\label{Con}  The contours near boundary of cylinder.}
\end{figure}

Consider an initial part of the sequence of contours when they appear near the boundary of cylinder. The second part of Proposition 1 claims that the boundary rotors $1,2,\dots$ flip their directions strictly sequentially. This order imposes a preferable direction for propagation of the contour sequence
$C_1,C_2,\dots$ against the initial direction of rotors in $\rho_0(v)$ (Fig.\ref{Con}).
When the sequence or contours makes a turn along the cylinder boundary, contour $\bar{C}_1$ becomes reachable for the rotor walk from the nearest contour $\bar{C}_{k-1}$. Then, the next contour $C_k$ starting from $\alpha_k$ passes a part of contour $\bar{C}_1$, boundary sites $1,2,\dots$ and returns to $\alpha_k$ via $\alpha_{k-1 }$ forming a non-contractible  loop around the cylinder. The orientation of the loop is still clockwise so that Proposition 1 is applicable to $C_k$. If $t_k$ is the moment when contour $C_k$  is closed at vertex $w_{t_k}$, then $\bar{t_k}$ is the moment when the rotor walk leaves the anticlockwise contour $\bar{C_k}$ at the same vertex. An important event happens in the time interval between $t_k$ and $\bar{t_k}$. To characterize it we prove the following statement.

\begin{figure}[!ht]
\includegraphics[width=40mm]{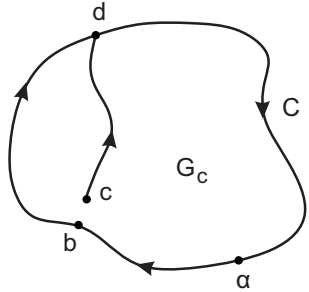}
\vspace{1mm}
\caption{\label{Prop} The clockwise contour with a spanning graph inside. Vertices $b$ and $c$ are neighbors on the graph. $\alpha$ is the initial chip position. }
\end{figure}

{\it Proposition 2}. Consider a planar graph $G$ and a rotor configuration representing a clockwise contour $C$ with a spanning graph $G_C$ inside. Let $\alpha, b,c,d$ be vertices such that $\alpha,b,d\in C$, $c\in G_C$, the edge $bc\in G$ does not belong to $G_C$, the vertex $b$ has only 3 neighbours in $G$, and $c$ is connected with $d$ by a directed path $p_{cd}\in G_C$. The initial chip position $\alpha$ is on a part of $C$ between $b$ and $d$ (Fig.\ref{Prop}). Starting at $\alpha$, let the rotor walk leaves anticlockwise contour $\bar{C}$ at some moment $t_{exit}$. Then, there exists a moment $t^{\star} < t_{exit}$ when a clockwise contour $bdcb$ occurs.

{\it Proof.} Fix temporarily the direction of the rotor at $b$ to $c$. By Proposition 1, rotors of the clockwise contour $\alpha,b,c,d,\alpha$ flip sequentially from $\alpha$ to $c$ via $d$. The first rotation of the rotor at $b$ is possible only after the jump of the chip from $c$ to $b$. But this is the moment $t^{\star}$ when contour $bdcb$ becomes closed if we return the rotor at $b$ to its initial place.

Returning to the sequence of contours near boundary depicted on Fig.\ref{Con}, we identify contour $C$ of the Proposition 2 with $C_k$, graph $G_C$ with the interior of $C_k$ together with boundary sites not belonging to $C_k$. The vertices $b$ and $c$ of the Proposition 2 are identified with the boundary vertices $1$ and $L$ and the path $p_{cd}$ with the path from $L$ to the contour $\bar{C}_{k-1}$. Then, by Propositions 1 and 2, there is a moment $t_k<t^{\star}<\bar{t_k}$ when the boundary contour $1,2,\dots,L,1$ becomes closed and a moment $t_k<t^{\star}+L<\bar{t_k}$ when the contour $1,L,L-1,\dots,1$
appears. The latter contour coincides with that in $\rho_0(v)$ and therefore, by the definition of excursion, the moment $t^{\star}+L$ is the end of the first excursion.

Three events turn out to be synchronized: the emergence of a non-contractible loop corresponding to a clockwise contour with label $\alpha_k$; the completing a turn around the cylinder by the sequence of clockwise contours; the end of the first excursion. This synchronization is preserved during the further evolution. Indeed, the rotors remaining on contours $\bar{C}_1,\bar{C}_2,\dots,\bar{C}_k$ constitute a forest attached to the boundary of cylinder. Then the rotor walk continuing from $w_{t_r}$ for $t>\bar{t}_r$ moves along this forest in the same way as the walk starting at $t=0$ moves along the original forest in $\rho_0(v)$. An essential difference is that the interiors of contours $\bar{C}_1,\bar{C}_2,\dots,\bar{C}_k$ are not available for new labels because no clockwise contours can be closed inside them. Therefore, the sequence of labels $\alpha_{k+1},\alpha_{k+2},\dots$ is disposed below the sequence $\alpha_{1},\alpha_{2},\dots,\alpha_{k}$ forming a next layer which propagates in the same direction until a label appears which corresponds to the next non-contractible loop. We come to a conclusion that a natural order for disposition of labels is a helix-like sequence on the surface of cylinder. The synchronization mentioned above implies that each turn of the helix corresponds to one excursion and to one label denoting a non-contractible contour.

The helical order of growth of the cluster of visited sites explains the mentioned non-uniformity of spanning trees generated by the rotor walk on the cylinder.Neverthelrss, the determination of the fractal dimension of the chemical path of the obtained tree remains an open problem.

\begin{figure}[!ht]
\includegraphics[width=85mm]{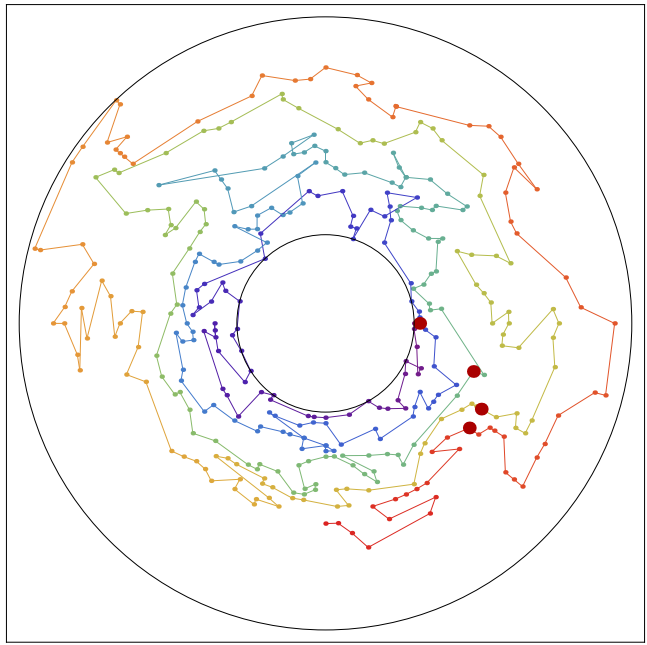}
\vspace{1mm}
\caption{\label{scr}  The labels on the equivalent annulus with the internal ring of length $L=100$. Bold points show the positions of labels corresponding to non-contractible loops.}
\end{figure}

A set of labels $\alpha_1,\dots,\alpha_m$, $m=369$, obtained for a particular random initial configuration $\rho_0(v)$ is shown in Fig.\ref{scr}.
The sequence $\alpha_1,\dots,\alpha_m$ demonstrates a helix structure which agrees with the construction described above and is expectedly random since $\rho_0(v)$ is random. We can characterize the positions of labels by variables $h(\alpha_i)$ and $\theta(\alpha_i)=2\pi k_i+\varphi_i$, where $h(\alpha_i)$ is the distance of $i$-th label from the top of cylinder, $k_i$ is the number of chip rotations around cylinder before reaching a given site and $\varphi_i$ is the polar angle. The helix-like structure implies that $h_i$ is asymptotically proportional to $\theta_i$.

The relation $h(\alpha_i)=b_i \theta(\alpha_i)$ can be examined for a large number of steps $i$ after averaging over large number of initial rotor configurations $\rho_0(v)$. Let $\hat{\alpha}_1,\hat{\alpha}_2$ be a subsequence of labels corresponding to the non-contractible loops selected from sequence of all labels $\alpha_1,\dots,\alpha_m$,$m\gg 1$. Four bold points in Fig.\ref{scr} show the positions of selected labels for four first rotations of labels around the cylinder. If the helix with increasing number of turns tends in average to a regular form, the average $\langle h(\hat{\alpha}_n)\rangle$ should meet the relation
\begin{equation}
\langle h(\hat{\alpha}_n)\rangle=b_L(n) 2\pi n
\end{equation}
where $b_L(n)$ tends to a constant $b_L=\lim_{n\rightarrow \infty} b_L(n)$ for every $L$.
\begin{figure}[!ht]
\includegraphics[width=100mm]{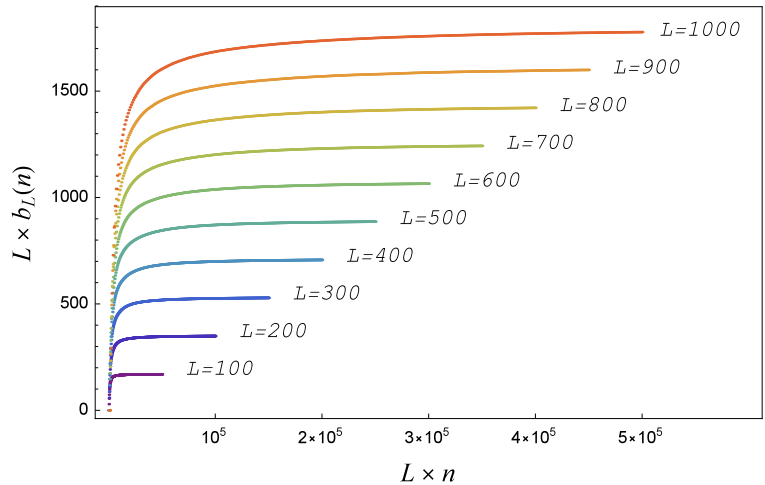}
\vspace{1mm}
\caption{\label{Fig} Coefficients $b_L(n)$ as functions of the number of excursions $1 \leq n \leq 500$ for cylinders of radius $L=100,\dots,1000$. }
\end{figure}
The functions $b_L(n)$ for different $L$ are shown in Fig.\ref{Fig}. For large $n$, we use the approximation
\begin{equation}
b_L(n)=b_L+\frac{A_L}{n}+\frac{B_L}{n^2}
\label{expansion}
\end{equation}
to find $b_L$.When $L$ increases, the values $b_L$ converge rapidly to the constant $b=\lim_{L\rightarrow \infty} b_L$ which is estimated as $b=1.80\pm0.05$.

The obtained value of the helix step $b$ can be compared with the estimation of the spiral step in \cite{PPP2}. In the case of planar lattice, we considered the average ratio of radius $r$ to angle $\theta$ $\langle r/\theta \rangle$ as a function of the number of labels which is similar to $b_L(n)$ in the present paper. In \cite{PPP2}, we were not able to declare a definite value of $\theta$ $\langle r/\theta \rangle$ in the limit of large numbers of the spiral rotations $n$ for a possible presence of logarithmic corrections $\sim \ln n$. In the case of cylindrical geometry, the expansion (\ref{expansion}) gives a perfect description of the large $n$ behavior and leads to the reliable value of $b$.

Despite the difference in the numerical estimations of the spiral and helical steps, the conceptual conclusions on the properties of the rotor-router walks are common in both geometries. In \cite{PPP2}, we have found that the number of visits of the origin $n_0$ depends on the number of spiral turns $n$ as
\begin{equation}
n_0=4n+O(1).
\end{equation}
Since one of four rotations of the initial rotor necessarily corresponds to the end of an excursion, we obtain the equivalence between the number of excursions and the number of spiral turns. The study in the present paper shows that this equivalence holds also for the helix turns.

\section*{Acknowledgments}
We thank Deepak Dhar for useful discussion and for drawing our attention to work \cite{singha}.
VBP thanks the RFBR for support by grant 16-02-00252.
VSP thanks the JINR program ``Ter-Antonyan - Smorodinsky'' and SCS MES RA project No 15YPR-1B0001.

\end{document}